\documentclass[12pt,a4paper]{revtex4}
\usepackage[utf8]{inputenc}
\usepackage{amsmath}
\usepackage{amsfonts}
\usepackage{amssymb}
\usepackage{graphicx}
\usepackage{indentfirst}
\usepackage{physics}
\usepackage[left=2cm,right=2cm,top=2cm,bottom=2cm]{geometry}

\usepackage{float}
\usepackage[caption = false]{subfig}
\usepackage[english]{babel}

\begin{document}

\title{Spontaneous symmetry breaking and Husimi Q-functions in extended Dicke model}

\author{S. S. Seidov$^{1,2}$}
\author{S. I. Mukhin$^{1}$}
\affiliation{$^{1}$Theoretical Physics and Quantum Technologies Department, NUST "MISIS", Moscow, Russia}
\affiliation{$^{2}$Vereshchagin Institute for High Pressure Physics, Russian Academy of Sciences, 108840 Troitsk, Moscow, Russia}

\begin{abstract}

We study the emergence of a parity breaking coherent photonic state of a photon mode coupled to finite array of  two-level systems, represented by pseudospins 1/2. The pseudospin-photon interaction is realised via a shift of the photonic oscillator equilibrium position by an amount linear in Cartesian component of the total pseudospin. We demonstrate analytically, that the instability is manifested in an upturn from concave to convex of the ground state energy  dependence on the total pseudospin component coupled to the photons. The perturbation, sufficient for parity breaking,  tends to zero in the ultrastrong limit of  light-matter coupling. We present phase diagram of finite pseudospin-photon system, that demonstrates this feature. Evolution of Husimi Q-functions of the pseudospin and photon subsystems, and of the pseudospin entropy, along different trajectories across the phase diagram is presented. 
\end{abstract}

\maketitle

\section{Introduction}
The Dicke model in cavity QED describes an ensemble of $N$ two-level systems (TLS) interacting with a single photonic mode \cite{Dicke, Garraway}. In thermodynamic limit there is a quantum phase transition described by parity symmetry breaking of the ground state wave function, though the symmetry is preserved by the Dicke Hamiltonian \cite{Brandes, Hepp}. The extended Dicke model with a quadratic term arises for various light--matter interaction models \cite{Saidi, Rabl,Rabl_2016, Mukhin}:
\begin{equation}
\hat H = \frac{\hat p^2 + \omega^2 \hat q^2}{2} + g \hat p \hat S_y - E_J \hat S_z + (1+\varepsilon)\frac{g^2}{2}\hat S_y^2.
\label{ED}
\end{equation}
Here the Plank constant $\hbar$ is put to one, $\hat S_i$ are total pseudospin operators, and photon momentum $\hat p$ and coordinate $\hat q$ operators are defined below, in Section II. When  $\varepsilon=0$, the full square with respect to $\hat p$ is completed and the spin--projection $S_y$ acts as a shift of the equilibrium position of the photonic oscillator. The phase transition properties for $\varepsilon \neq 0$ were studied in \cite{Rabl, Rabl_2016}. It was shown, that for $\varepsilon<0$ there is a transition into superradiant state, while a subradiant state realizes for $\varepsilon>0$ and in narrow region of $\varepsilon<0$ domain. In this work we focus on the $\varepsilon - g^2$ phase diagram for finite $N$ system, in the  limit $g^2\rightarrow \infty$ .  

The case of $\varepsilon=0$ was investigated previously \cite{Mukhin}, where a first order quantum phase transition (QPT) was found in thermodynamic limit $N\gg 1$ using rotating Holstein--Primakoff transformation. The parity symmetry of the ground state wave function gets broken as a result of formation of the superradiant photonic condensate. However, the existence of a phase transition in the Dicke model with quadratic term was subjected to debate \cite{Rzazewski, Keeling, Stokes, Bialynicki-Birula}. Besides, it is obvious, that for finite total spin the parity symmetry of the Hamiltonian is always preserved \cite{Ashhab, IMAI20183333, Shen_2017, Shen_2020, Hwang}. 
This contradiction is studied in the present paper. Finite $N$ system with the Hamiltonian (\ref{ED}) gets spontaneous symmetry breaking along the $g$-axis at $\varepsilon=0$  in the limit $g^2\rightarrow \infty$. Though the Hamiltonian symmetry is formally conserved for any finite values of parameters, the parity symmetric phase becomes unstable to a formation of a shift of the equilibrium position of momentum $p$ of the cavity mode. We prove by a straightforward analytical derivation, that the perturbation, sufficient for parity breaking via an emergence of photonic  shift, tends to zero in the ultrastrong limit of  light-matter coupling. Simultaneously, the latter perturbation causes an upturn from concave to convex of the dependence of the ground state energy on the total spin--projection $S_y$ involved in light-matter coupling in Hamiltonian (\ref{ED}). 

The plan of the paper is as follows. First, we briefly dwell on the superradiant and subradiant quantum phase transitions discussed in the earlier works \cite{Brandes, Rabl_2016, Rabl}. Then focus is made on the limit $g^2\rightarrow \infty$ in finite $N$ TLS, where the Born--Oppenheimer potentials could be used to illustrate the difference between various types of phases.  An analysis of spontaneous symmetry breaking is then made using idea \cite{Wezel, Aron} of non--commuting limits of ultrastrong coupling and of vanishing non-zero symmetry breaking term (finite superradiant density). For this purpose, a perturbation theory is used in the strong coupling limit, to obtain ground state energy as a function of the spin--projection $S_y$, involved in light-matter coupling, and of a finite photonic equilibrium shift $\alpha$, that breaks the symmetry of the ground state. An evolution of the Husimi Q-functions of the photon and spin subsystems along the different pathways on the phase diagram is presented, that directly exhibits the changes in the behaviour of the subsystems under an increase of spin-photon coupling strength $g$ at different values of parameter $\varepsilon$.  In the last section, we demonstrate numerically  an occurrence of the phase transition in the strong coupling limit in the case $\varepsilon=0$ using broken symmetry trial wave-function method. 

\section{The model}
We consider an extended Dicke Hamiltonian of an ensemble of $N$ two-level systems interacting with bosonic mode: 
\begin{equation}
\hat H = \omega \hat a^\dagger \hat a + ig \sqrt\frac{\omega}{2}(\hat a^\dagger - \hat a)\sum_j^N \hat\sigma^y_j - E_J \sum_j^N \hat\sigma^z_j + (1+\varepsilon) \frac{g^2}{2} \sum_{jk}^N \hat\sigma^y_j \hat\sigma^y_k.
\label{D}
\end{equation} 
A single two-level system is described by spin $1/2$ with projections operators $\hat\sigma^{x,y,z}_j$. The $\propto g^2\varepsilon$ term arises from dipole-dipole interaction \cite{Rabl, Rabl_2016}. Depending on $\varepsilon$, one obtains the usual Dicke model ($\varepsilon = -1$)  for strong ferroelectric inter-dipole interaction  (i.e. inter-spin interaction in the spin-$1/2$ representation of the two-level systems),  or an extended Dicke model for repulsive inter-dipole interaction ($\varepsilon >0$). The  Hamiltonian with $\varepsilon = 0$ is obtained for Josephson junctions interacting merely via the photonic mode \cite{Saidi,Mukhin}, through a gauge invariant shift of Cooper pairs phases differences across the junctions. 

Introducing collective spin operators $\hat S_{x,y,z} = \sum_j \hat \sigma^{x,y,z}_j$ we rewrite (\ref{D}) in the form:
\begin{equation}\label{eq:H}
\hat H = \omega \hat a^\dagger \hat a + ig \sqrt\frac{\omega}{2}(\hat a^\dagger - \hat a)\hat S_y - E_J \hat S_z + (1 +\varepsilon) \frac{g^2}{2}\hat S_y^2.
\end{equation}
For qualitative investigation it is convenient to switch to momentum and coordinate operators constituting harmonic oscillator of the photon mode:
\begin{align}
&\hat q = \sqrt\frac{1}{2\omega}(\hat a^\dagger + \hat a)\\
&\hat p = i\sqrt\frac{\omega}{2}(\hat a^\dagger - \hat a).
\end{align} 
The Hamiltonian then is expressed as: 
\begin{equation}
\hat H = \frac{\hat p^2 + \omega^2 \hat q^2}{2} + g \hat p \hat S_y - E_J \hat S_z + (1+\varepsilon)\frac{g^2}{2}\hat S_y^2.
\label{eq:Hr}
\end{equation}
One can see that a full square can be completed:
\begin{equation}
\hat H = \frac{\omega^2 \hat q^2}{2} + \frac{1}{2}(\hat p + g \hat S_y)^2 - E_J \hat S_z + \frac{\varepsilon g^2}{2} \hat S_y^2\equiv \frac{\omega^2 \hat q^2}{2} +\hat{U},\label{eq:H2}
\end{equation}
where $\hat{U}$ plays the role of adiabatic potential $U(p, \sigma_y)$ in the ultrastrong coupling limit $E_J/g^2\rightarrow 0$, as is discussed below. When $\varepsilon = 0$ and $E_J = 0$, the $\hat S_y$ can be treated as a gauge field, and the Hamiltonian becomes gauge invariant because $g \hat S_y$ only shifts equilibrium position of the oscillator, but does not effect its energy. 

\subsection{Parity conservation and spontaneous symmetry breaking}
The Hamiltonian (\ref{eq:H}) commutes with the parity operator
\begin{equation}\label{eq:Parity}
\hat \Pi = \exp\qty{i \pi \qty(\hat a^\dagger \hat a + S + \hat S_z)},
\end{equation}
so there is a corresponding conservation law. Neighbouring energy levels in the spectrum have different parity. The operator
\begin{equation}
\hat{\mathcal{N}} = \hat a^\dagger \hat a + S + \hat S_z
\end{equation}
is the excitations number operator, here $S$ is the total spin value. Each term in the Hamiltonian changes the number of excitations by 2 because it is either quadratic in $\hat p$, $\hat q$ and $\hat S_y$ or is their product $\hat p \hat S_y$, so the parity stays unchanged. 

The Hamiltonian commutes with $\hat \Pi$ for any values of parameters and formally the symmetry will be always preserved. However, there is still a possibility for a phase transition in the strong coupling limit. First of all, as coupling increases, the energy gap between states with different parity tends to zero, meaning they will mix and the conservation law could be broken by a small external perturbation. A clear sign of spontaneous symmetry breaking is non-commuting limits of infinite coupling and infinitesimal symmetry breaking terms. We consider this in detail in the next sections. Parity symmetry breaking in the superradiant state becomes evident if one allows for the photonic condensate, e.g. in the form of a c-number shift of the $\hat{p}$ operator:
\begin{equation}\label{eq:shift}
\hat{p} \rightarrow \hat{p} + \ev{\hat{p}} ,
\end{equation}
 that causes appearance of the terms linear in $\hat{p}$ in the Hamiltonian (\ref{eq:Hr}), thus, making it not commuting with the parity operator $\hat \Pi$ \cite{Brandes, Mukhin}.

\section{Superradiant and subradiant phases}
In previous works \cite{Rabl, Brandes} it was shown, that depending on the value of $\varepsilon$ in the Hamiltonian (\ref{eq:H2}), the system acquires different types of phases in the ground state. The distinction manifests itself in the macroscopic observables, such as averages of photons number and pseudospin projection. 

To illustrate the different phases we calculate Born--Oppenheimer potential of the Hamiltonian in strong coupling limit (i.e. $E_J/g^2\rightarrow 0$). It is a set of parabolas:
\begin{equation}\label{eq:E_infty}
U_0(p,\sigma_y)=\frac{1}{2}(p + g \sigma_y)^2 + \frac{\varepsilon g^2}{2}\sigma_y^2, 
\end{equation}
centred at $p=-g\sigma_y $, with gaps opened at their crossings due to the term $-E_J \hat S_z$  in (\ref{eq:H2}), $\sigma_y$ is an eigenvalue of $\hat S_y$. Hence, the multi--well potential possesses $2S+1$ minima \cite{Cohn, Rabl}. Depending on the sign of $\varepsilon$, different potential wells will be the lowest ones: at $p = \pm g S$ for negative $\varepsilon$ , and at $p = \pm g/2$ or  $p =0$ for positive $\varepsilon$. In the latter case the minimum is at $p=0$ for integer spin and at $p=\pm g/2$ for half--integer.

Due to parity symmetry the potential $U_0(p,\sigma_y)$ is also symmetric, which leads to a degeneracy between the states in the potential wells symmetric with respect to $p=0$. Finite $E_J$ term lifts the degeneracy, and the energy levels of previously degenerate states split.

\subsection{Superradiant phase}
\begin{figure}[h!!]
\center\includegraphics[scale=0.8]{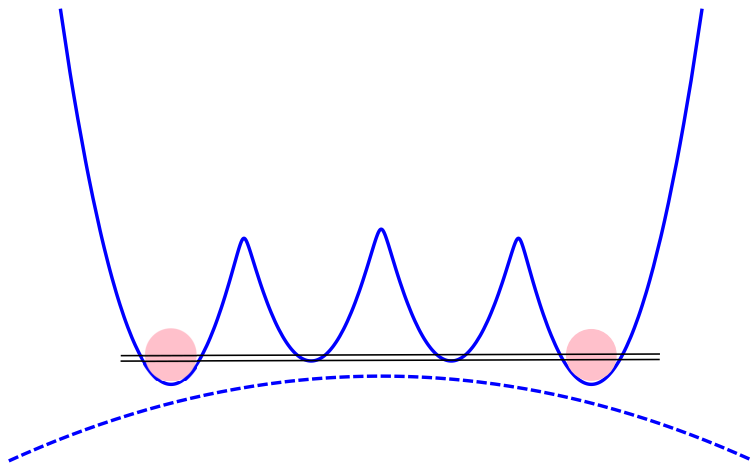}
\caption{Born--Oppenheimer potential $U(p,\sigma_y)$ of the system in superradiant phase. Thin black lines schematically illustrate energy level splitting between two degenerate states, see Eq. (\ref{eq:Psi_infty}), in symmetric potential wells corresponding to spin projection $\sigma_y = \pm S$. The dashed line underlines a convex curvature of $U$ with respect to $p$.}
\label{fig:BO_pot_down}
\end{figure}

The superradiant quantum phase transition occurs when $\varepsilon < 0$. In the latter case the critical coupling is \cite{Rabl}: $g_c \sim \sqrt\frac{E_J}{-\varepsilon N}$, and the average photonic oscillator momentum and spin projection on $y$--axis after the transition are equal, respectively, to ones of those: $\ev{\hat p} \sim -g\ev{\hat{S}_y} \sim \pm gN/2$. The Born--Oppenheimer potential has lowest wells at its ends: $p=\pm gN/2$, as is illustrated in fig. \ref{fig:BO_pot_down}. So, at the phase transition, the system breaks the symmetry and finds itself in one of the corresponding wells with non-zero averages of $\ev{\hat p}$ and $\ev{\hat{S}_y}$. 

\subsection{Subradiant phase}
\begin{figure}[h!!]
\subfloat[$\varepsilon >0$, integer $S$]{\includegraphics[width=0.4\linewidth]{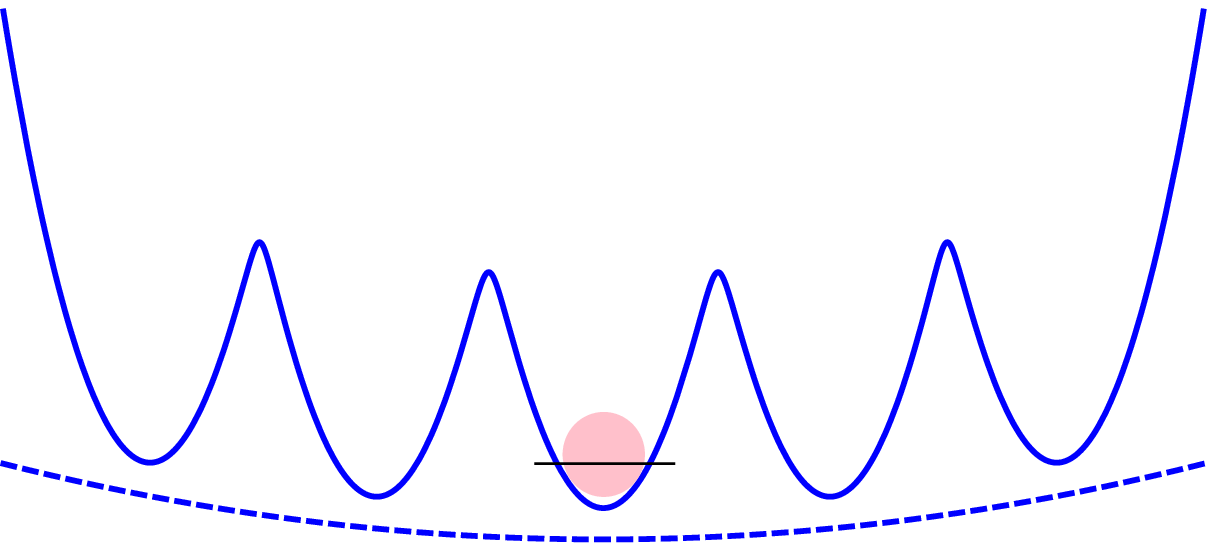}}\hspace{1cm}
\subfloat[$\varepsilon >0$, half--integer $S$]{\includegraphics[width=0.4\linewidth]{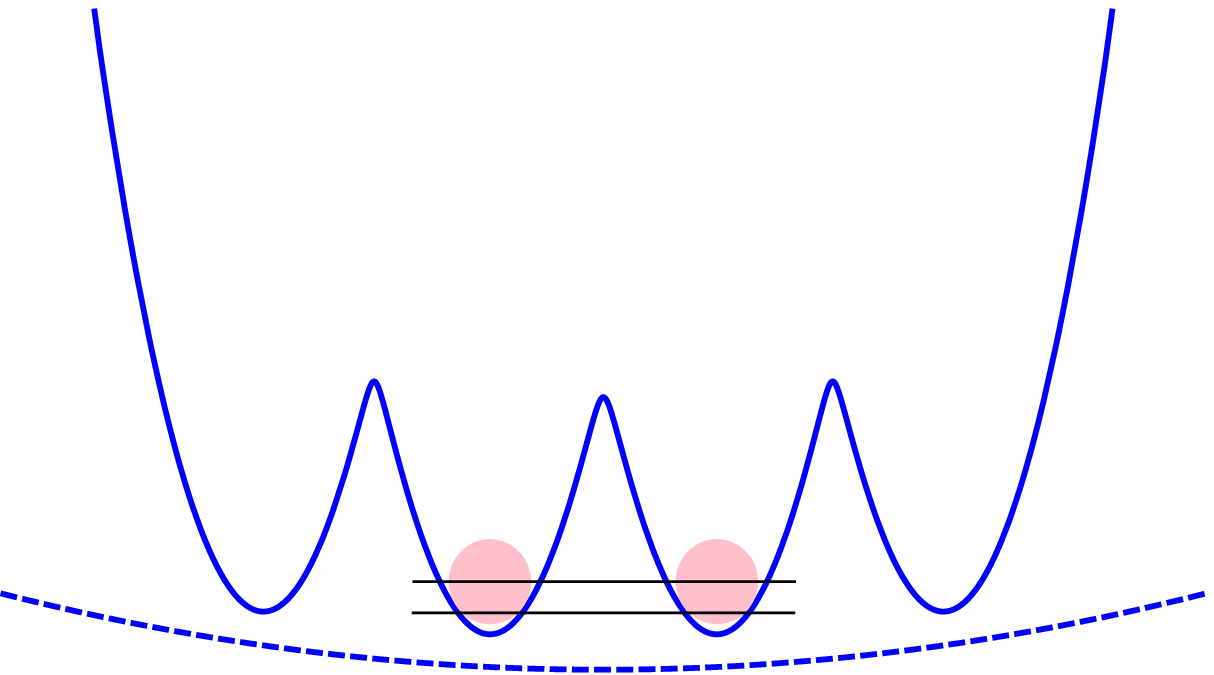}}
\caption{Born--Oppenheimer potential $U(p,\sigma_y)$ of the system in subradiant phase. Thin black lines schematically illustrate energy level splitting between two barely degenerate states in the symmetrically positioned potential wells, corresponding to spin projection $\sigma_y = \pm 1/2$, see Eq. (\ref{eq:Psi_infty}), or a single non-split level with $\sigma_y = 0$ for integer $S$.}
\label{fig:BO_pot_up}
\end{figure}
In subradiant phase (i.e. for $\varepsilon > 0$) the lowest wells of the Born--Oppenheimer potential are the central ones, see  fig. \ref{fig:BO_pot_up}. Hence, the projection of pseudospin $S_y$ on $y$--axis in the lowest potential wells is minimal, being $\pm1/2$ for odd $N$, and $0$ for even $N$.
The subradiant phase occurs as a cross-over \cite{Rabl}, see Fig. \ref{fig:PDG}. Corresponding ground state wave function does not depend on the system size $N$, because in this phase the Born-Oppenheimer potential possesses minima of energy at the pseudospin projections $S_y$: $S_y=\pm1/2$ or $S_y=0$. Hence, the wave-function of the photonic oscillator is concentrated at the coordinates: $p=\pm g/2$ or $p=0$ respectively, that do not depend on the total spin $S$.

\subsection{Intermediate case: $\varepsilon=0$}
\begin{figure}[h!!]
\center\includegraphics[scale=0.8]{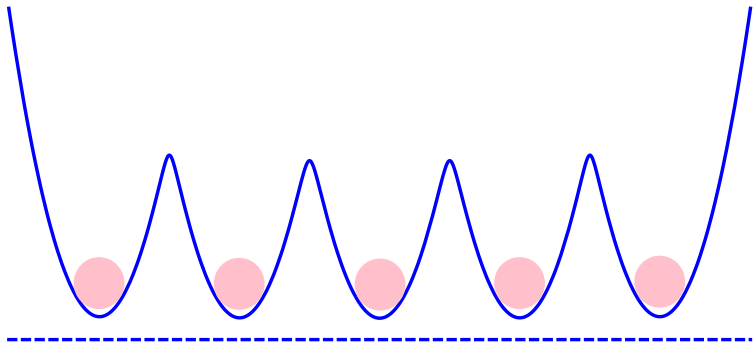}
\caption{Born--Oppenheimer effective potential $U(p,\sigma_y)$ of the photon field oscillator at strong coupling $g$ for $\varepsilon=0$. All wells are on the same energy level, so the system is in highly delocalized state.}
\label{fig:BO_pot_flat}
\end{figure}

The Hamiltonian with $\varepsilon=0$ arises in the description of an array of Josephson junctions coupled to microwave cavity via gauge--invariant shift of the junctions phases \cite{Saidi, Mukhin}. The Hamiltonian expressed via photon oscillator coordinate and momentum is:
\begin{equation}
H = \frac{\omega^2 \hat q^2}{2} + \frac{1}{2}\qty(\hat p + g \hat S_y)^2 - E_J \hat S_z.
\label{Dickeini}
\end{equation}
One can clearly see, that $\hat S_y$ acts as a gauge field. For strong coupling $g$, when the $\hat S_z$ term is negligible, the Hamiltonian is invariant with respect to change of the projection $\sigma_y$. This means that in Born--Oppenheimer potential all wells share the same energy level, see fig. \ref{fig:BO_pot_flat}. 
However, any small $E_J$ will destroy the degeneracy, making preferable the spin state with maximal total spin $S$  and  with the minimal projection on $y$--axis \cite{Rabl}. At first glance, this will lead to subradiant phase transition, as for $\varepsilon > 0$, but it is not the case. The transition has its own unique properties which we discuss below (see also \cite{Mukhin}).  

\subsection{Spontaneous symmetry breaking}
The system's wave function $\psi$ depends on coupling constant $g$ in the Hamiltonian (\ref{Dickeini}), but a parity of the state described by the eigenvalue of the operator $\hat \Pi$ in Eq. (\ref{eq:Parity}) is preserved by the Hamiltonian for any $g, N$. Now, we add an infinitesimal symmetry breaking field proportional to some small parameter $\alpha\rightarrow 0$. Then, if the limits  $g \rightarrow \infty$ and $\alpha \rightarrow 0$ do not commute i.e.:
\begin{equation}\label{eq:lims}
\lim_{g \rightarrow \infty} \lim_{\alpha \rightarrow 0} \psi \neq \lim_{\alpha \rightarrow 0} \lim_{g \rightarrow \infty} \psi ,
\end{equation}
the system will undergo a spontaneous symmetry breaking in $g \rightarrow \infty$ limit \cite{Wezel}. 

\subsection{Superradiant instability at $\varepsilon=0$}
To study this phenomenon of not commuting limits analytically in the strong coupling limit of $g$, we add small symmetry breaking term $\alpha \hat p$  to the Hamiltonian and treat $-E_J \hat S_z$ term as a perturbation $\hat H_{1}$ :
\begin{equation}\label{eq:H_shift}
\hat H_{0,\alpha} +\hat H_{1}= \frac{\omega^2 \hat q^2}{2} + \frac{1}{2}\qty(\hat p + g \hat S_y)^2 + \alpha \hat p -E_J \hat S_z\,.
\end{equation} 
The energy spectrum (counted from zero-point energy $\omega/2$) and wave functions of unperturbed Hamiltonian $\hat H_{0,\alpha}$ are:
\begin{equation}\label{eq:E_ab}
E^{(0)}_{\alpha,n} = \omega n - \frac{\alpha^2}{2}-g \alpha \sigma_y 
\end{equation}
\begin{equation}\label{eq:Psi_ab}
\ket{\psi^{\alpha}_{n,\sigma_y}} = \ket{n_{g \sigma_y + \alpha}, \sigma_y}\equiv \ket{n_{g \sigma_y + \alpha}}\ket{\sigma_y},
\end{equation}
\noindent
where index $\sigma_y$ defines total spin projection and $\ket{n_{g\sigma_y + \alpha}}$ is a photon oscillator $n$-th excited bare eigenstate, with momentum minimum shifted by an amount of $-(g\sigma_y + \alpha$). Such a state is defined as:
\begin{equation}
\ket{n_{\ev{p}}} = e^{i\ev{p} \hat q}\ket{n},
\end{equation}
where $\ket{n}$ is the $n$--th excited state of the harmonic quantum oscillator. Now, to second order in $E_J$,  perturbation theory gives the following correction to the energy of the ground state $E^{(0)}_{\alpha,0}$ (compare \cite{Rabl_2016}):
\begin{equation}\label{second_order}
\begin{aligned}
&E^{(2)}_{\alpha,0} = -\sum_{n=1}^{\infty}\left\{\frac{|\bra{n_{g (\sigma_y-1) + \alpha}, \sigma_y-1} \hat H_{1}\ket{0_{g \sigma_y + \alpha}, \sigma_y}|^2}{\omega n+g\alpha} \right. \\
&\left. +\frac{|\bra{n_{g (\sigma_y+1) + \alpha}, \sigma_y+1} \hat H_{1}\ket{0_{g\sigma_y + \alpha}, \sigma_y}|^2}{\omega n-g\alpha}\right\}\\
&\approx\frac{E_J^2}{2g^2}\left\{[\sigma_y^2-S(S+1)]\left(1-\frac{\gamma^2e^{-\gamma^2} \sinh{(2\delta\ln{\gamma})}}{\delta}\right)- 2\frac{\sigma_y}{\delta}\gamma^2e^{-\gamma^2}\sinh^2{(2\delta\ln{\gamma})}\right\}
\end{aligned}
\end{equation}
\noindent
where: $\gamma^2=g^2/\omega$ and $\delta=g\alpha/\omega$. First, we take limits in the order of the r.h.s. of Eq. (\ref{eq:lims}). For $\gamma \gg 1$ and $\delta<\gamma^2$ terms $\sim e^{-\gamma^2}$ in (\ref{second_order}) can be neglected and using (\ref{eq:E_ab}) one finds:
\begin{equation}\label{E2d}
E_0=E^{(0)}_{\alpha,0} +E^{(2)}_{\alpha,0}= - \frac{\alpha^2}{2}-g \alpha \sigma_y +\frac{E_J^2}{2g^2}\left(\sigma_y^2-S(S+1)\right).
\end{equation}
\noindent Now, minimize $E_0$ with respect to $\sigma_y$ :
\begin{equation}\label{alpham}
\frac{\partial E_0 }{\partial\sigma }=0;\; \Rightarrow \alpha_m=E_J^2\sigma_y/g^3.
\end{equation}
\noindent
Then, substitute the resulting expression for $\alpha_m$ back into (\ref{E2d}) and find the minimal energy as function of $\sigma_y(\alpha)$:
\begin{equation}\label{Econvex}
E_{0,rhs}=E^{(0)}_{\alpha_m,0} +E^{(2)}_{\alpha,0}= - \frac{E_J^2}{2g^2} \left(\sigma_y^2+S(S+1)\right)+O(E_J^4/g^{6}).
\end{equation}
\noindent On the other hand, if we take limits in the order of the l.h.s. of Eq. (\ref{eq:lims}), i.e. the limit $\alpha\rightarrow 0$ is implemented first, one finds \cite{Rabl_2016}:

\begin{equation}\label{Econcave}
E_{0,lhs}=E^{(0)}_{\alpha=0,0} +E^{(2)}_{\alpha=0,0}= \frac{E_J^2}{2g^2} \left(\sigma_y^2-S(S+1)\right).
\end{equation}
\noindent Remarkably,  the sign of the curvature of the ground state energy $E_0$ as function of pseudo-spin projection $\sigma_y$ is different in (\ref{Econvex}) and (\ref{Econcave}). Hence, indeed, the limits in (\ref{eq:lims}) do not commute. Namely, the  limit of conserved parity $\alpha=0$ taken prior to $g\rightarrow \infty$, produces concave dependence of  $E_0$ on $\sigma_y$, see (\ref{Econcave}), as is the case in subradiant configuration of the Born--Oppenheimer effective potential in Fig. \ref{fig:BO_pot_up}. On the other hand, in the limit $g\rightarrow \infty$, and $\alpha=E_J^2S/g^3\rightarrow 0$, i.e. with infinitesimal parity breaking term $\alpha \hat p$ in the Hamiltonian (\ref{eq:H_shift}), the ground state energy $E_0$ is convex function of pseudo-spin projection $\sigma_y$, see (\ref{Econvex}), as is the case in the superradiant configuration of the Born--Oppenheimer effective potential in Fig.\ref{fig:BO_pot_down}. Thus, a transition into parity broken, "dipole ordered" state with $\sigma_y=\pm S = \pm N/2$ takes place in this case. This conclusion is in accord with the result found previously \cite{Mukhin}, that the "dipole ordered" state in the $\epsilon=0$ model occurs under condition $g\geq g_c =\sqrt{4E_JS}$ in the thermodynamic limit $N \gg 1$, since in this case $g_c=\sqrt{2E_JN}\rightarrow \infty$ allowing for $S=N/2$. 

A few observations are in order here. First, we emphasise, that both dependencies  $E_{0,rhs}$ and   $E_{0,lhs}$  in (\ref{Econvex}) and (\ref{Econcave}) are obtained here for the case $\varepsilon=0$, though, they resemble Figures.\ref{fig:BO_pot_down} and \ref{fig:BO_pot_up} for the  cases $\varepsilon<0$ and $\varepsilon>0$ respectively. Second, our result in Eq. (\ref{Econvex}) indicates that in the parity broken ground state 
the shift of photonic equilibrium $\ev{\hat p}$ in the extended Dicke Hamiltonian Eq. (\ref{eq:H_shift}) is:
\begin{equation}\label{avp}
\ev{\hat p + g \hat S_y+\alpha}=0,\; \Rightarrow \ev{\hat p}=\mp gS-\alpha_m \sim \mp gS+O\qty(\frac{E_J^2S}{g^3})\,, 
\end{equation}
\noindent  allowing for  the value of $\alpha_m$ from Eq. (\ref{alpham}). The shift of photonic equilibrium $\ev{\hat p}$ in (\ref{avp}) is in accord with the result obtained in \cite{Mukhin} for the amplitude of photonic superradiant condensate  in the limit $N \gg 1$. On the other hand, inequalities: $g^2\gg \omega$ and ${E_J^2S}/{g^4}<<1$
leave space for occurrence of the symmetry breaking transition already in the case  $N\sim 1$, though, it would be not a thermodynamic phase transition in a strict sense (compare \cite{Ashhab, IMAI20183333}). Namely, the number of photons, indeed, follows 'thermodynamic limit' as long as the coupling strength goes to infinity: $\ev{n}\sim\ev{\hat p}^2 \sim g^2\rightarrow \infty$,  while the number of TLS remains finite: $N\sim 1$.

\subsection{Ground state wave function}
We now focus on the ground state wave function. The energy (\ref{eq:E_ab}) is minimal for $n=0$ and $\sigma_y = S$ if $g,\alpha > 0$. Then, the ground state wave function is:
\begin{equation}
\ket{\psi^{\alpha}_{0,S}} = \ket{0_{gS + \alpha}, S}.
\end{equation}
From Born--Oppenheimer potential point of view this may be thought of as if the symmetry breaking term has added a slope to the initially "horizontal" bottom line in Fig. \ref{fig:BO_pot_flat}, thus, driving the system to the well at the edge, corresponding to $\sigma_y = S$. Since the lowest potential well is now positioned at $p=-gS$, in the limit of $g \rightarrow \infty$ the system occupies state far from the central potential well ($p=0$) in Fig.\ref{fig:BO_pot_flat}. Simultaneously the parity operator (\ref{eq:Parity}) does not commute with Hamiltonian (\ref{eq:H_shift}).

However, when $\alpha= 0$ the Hamiltonian eigenfunctions must simultaneously be the eigenfunctions of the parity operator $\hat\Pi$. The wave functions (\ref{eq:Psi_ab}) do not satisfy this condition and the symmetric wave functions are \cite{Bastarrachea_lattice} :
\begin{equation}\label{eq:Psi_infty}
\ket{\psi_{n,\sigma_y}^{0}} = \frac{1}{\sqrt{2(1+\delta_{\sigma_y,0})}}\qty(\ket{n_{g\sigma_y}, \sigma_y} + (-1)^n\ket{n_{-g\sigma_y}, -\sigma_y}),
\end{equation}
where $\delta_{\sigma_y,0}= \braket{0}{\sigma_y}$.
The ground state energy in this case does not depend on $\sigma_y$, so, the ground state is degenerate and symmetric. It means, that the state of the system  is a superposition of states symmetric with respect to $p=0$ well.

Let us start from the state $\ket{\psi^{\alpha}_{0,S}}$. If we first take the limit of zero symmetry breaking, i.e. $\alpha \rightarrow 0$, the wave function will restore the superposition with $\sigma_y=-S$ state via tunnelling to symmetric potential well and acquiring the form (\ref{eq:Psi_infty}). Now, it will also remain symmetric in the limit of $g \rightarrow \infty$.  

Alternatively, if we take the limits in the different order, i.e the limit $g \rightarrow \infty$ will be taken first, then the system will not be able to restore the superposition state (\ref{eq:Psi_infty}) after $\alpha \rightarrow 0$ limit, because the symmetric counterpart of the system's state will be localized infinitely far at $p\sim -gS$. One can see, that depending on the order of the limits, the system reaches the different states: either symmetric or not. This means that in the infinite coupling limit (or  in the zero photonic mode limit $\omega\rightarrow 0$ \cite{Ashhab}) the symmetric state is unstable and a spontaneous symmetry breaking will occur (compare \cite{IMAI20183333}). 

\subsection{Superradiant instability at finite $|\varepsilon|$}
\begin{figure}[h!!]
\center\includegraphics[scale=0.8]{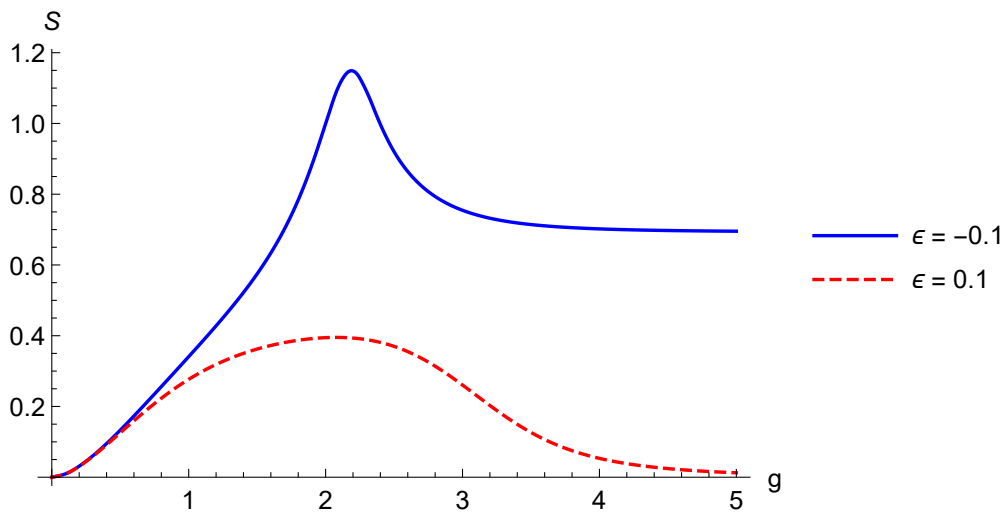}
\caption{Dependence of spin subsystem entropy on coupling constant for positive and negative $\varepsilon$. The plot is made for total spin $S = 2$.}
\label{fig:Entropy}
\end{figure}

\begin{figure}[h!!]
\center\includegraphics[scale=0.75]{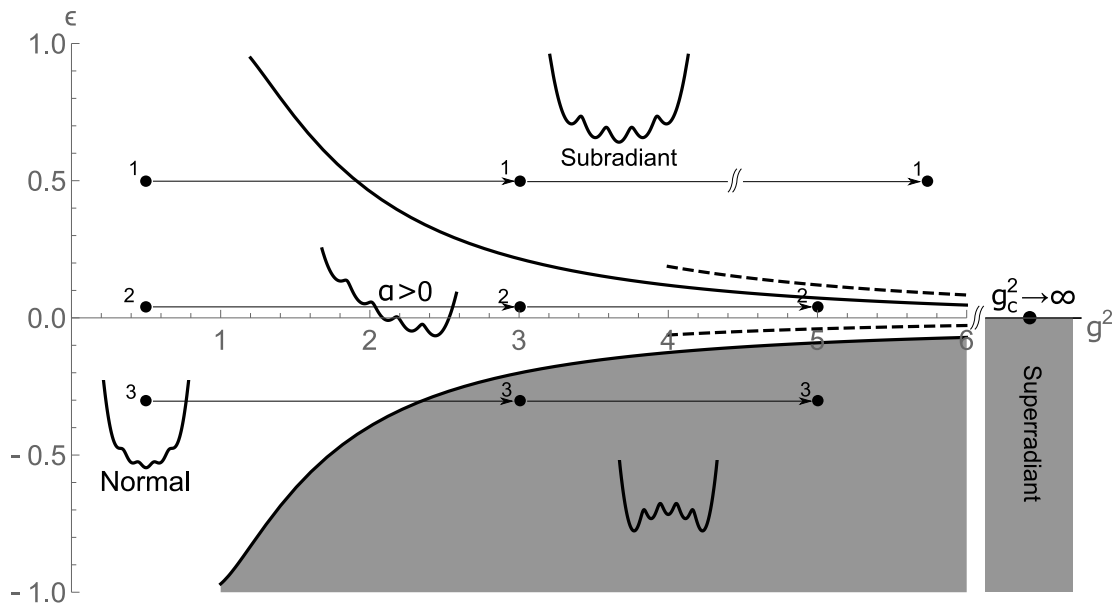}
\caption{The phase diagram $g^2-\varepsilon$ at finite $S$. The $g^2$-axis contains superradiant state at infinity, as is considered in the paper. Solid lines are obtained numerically via finding the maximum of spin subsystem entropy as function of coupling constant $g$. Dashed lines are analytical results for the instability boundaries (\ref{alpde}) and (\ref{convexee}) at large coupling constants. Arrows connect points on the three different routes parallel to $g^2$-axis, for which Husimi Q-functions are presented in Fig. \ref{fig:HQ}.}
\label{fig:PDG}
\end{figure}

\begin{figure}[h!!]
\flushleft\subfloat[a]{\includegraphics[scale=0.45]{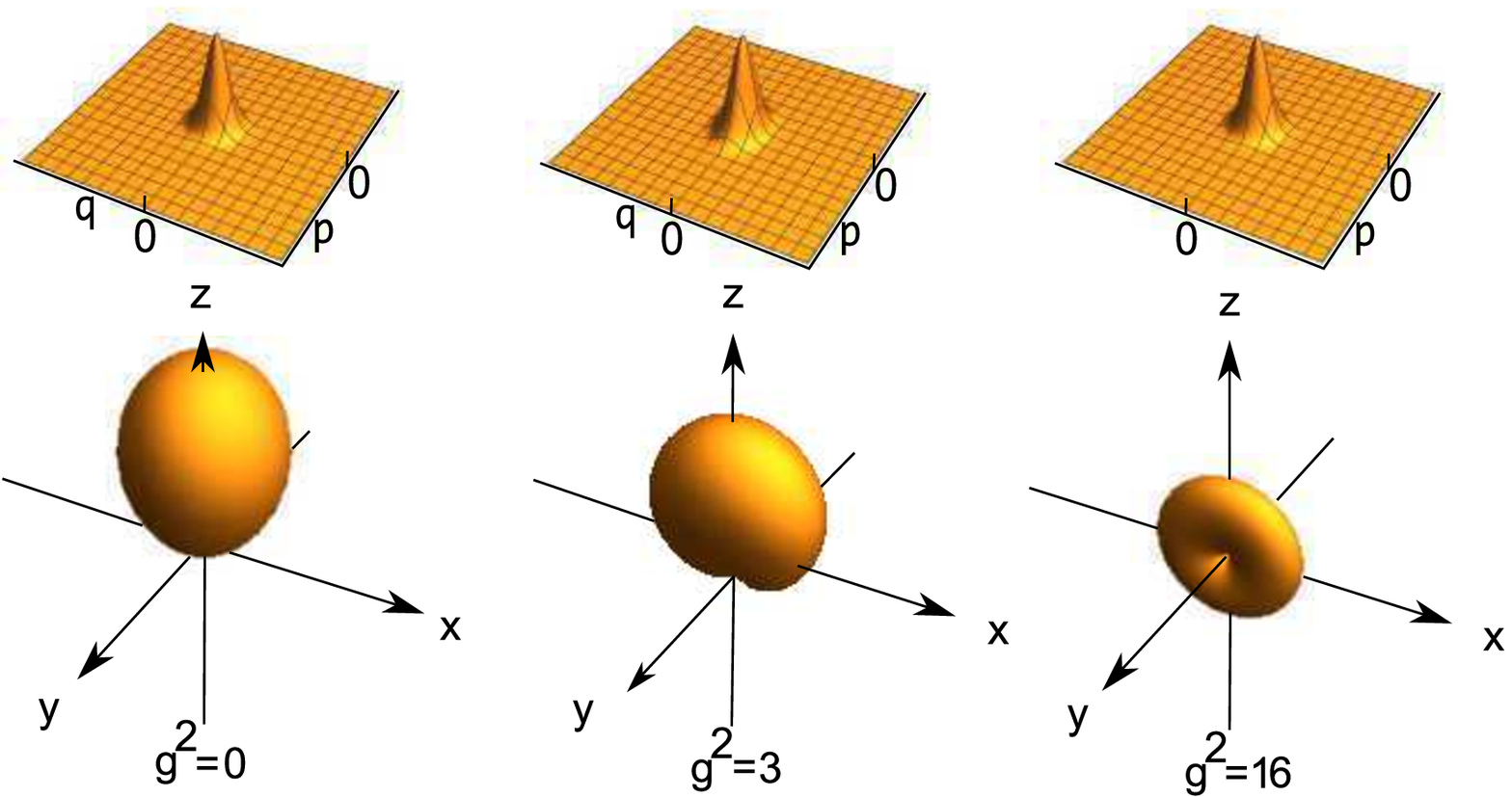}}\hspace{0.7cm}
\subfloat[b]{\includegraphics[scale=0.45]{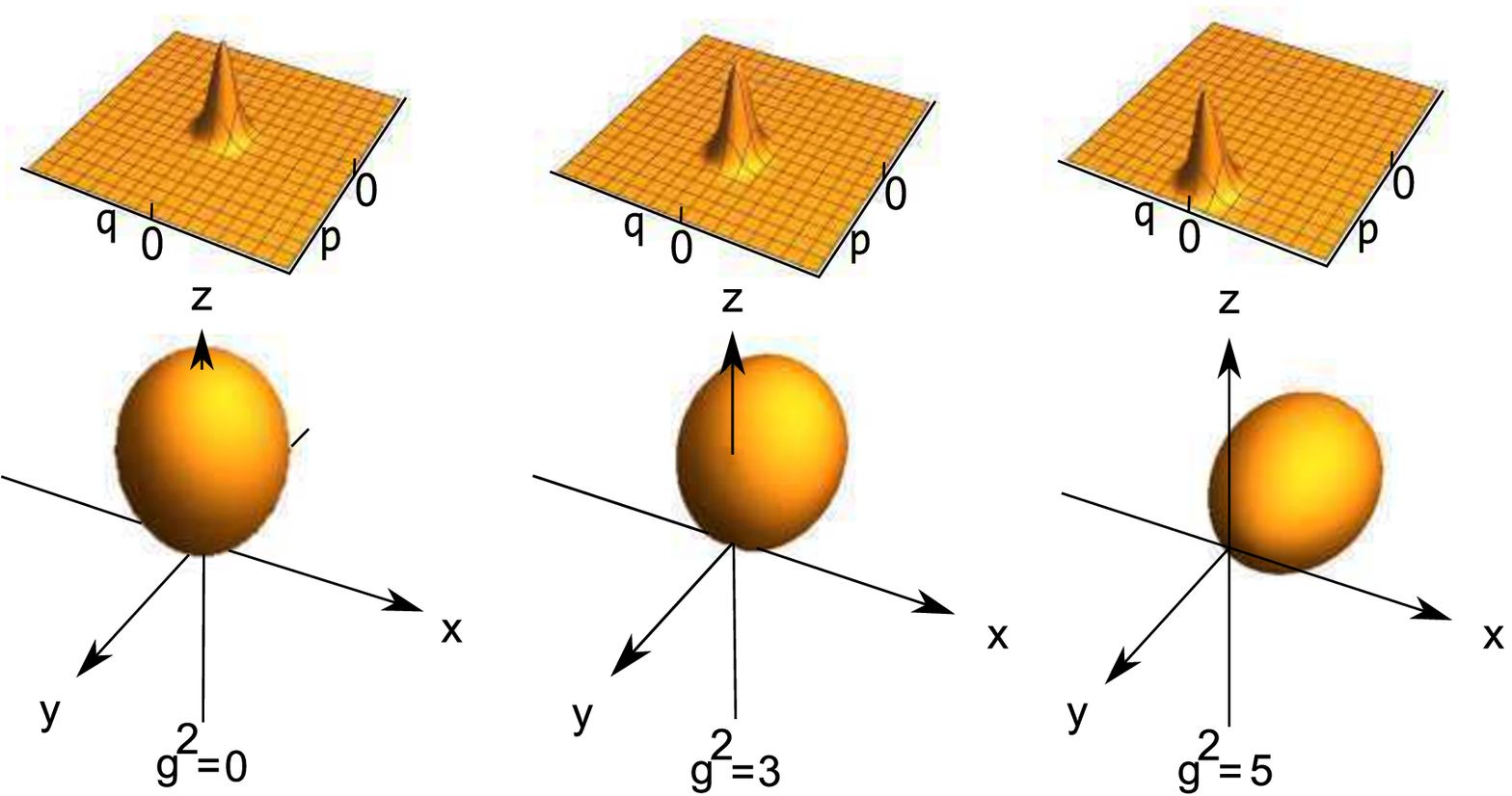}}\\
\flushleft\subfloat[c]{\includegraphics[scale=0.45]{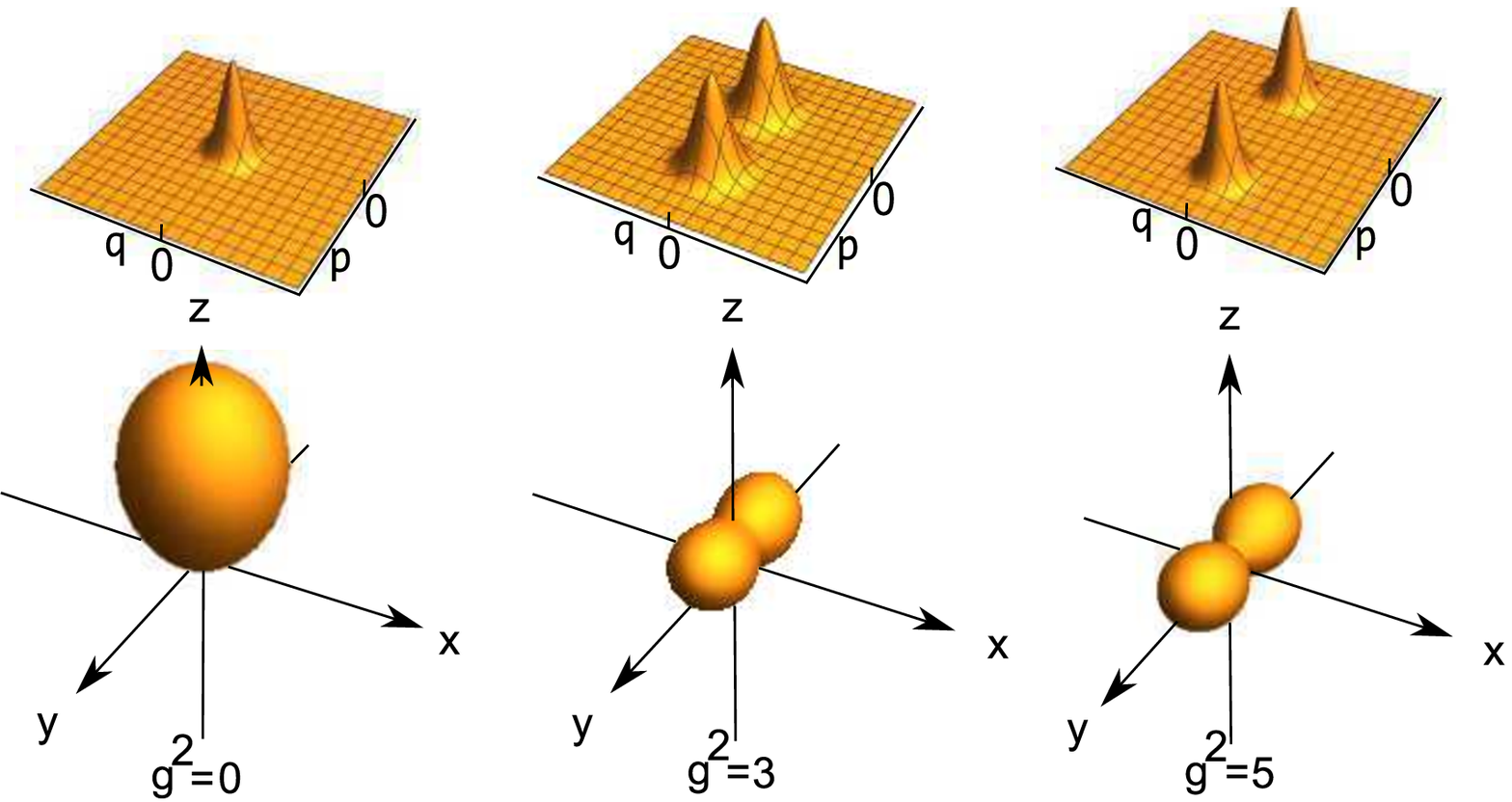}}\hspace{1.0cm}
\subfloat[d]{\includegraphics[scale=0.2]{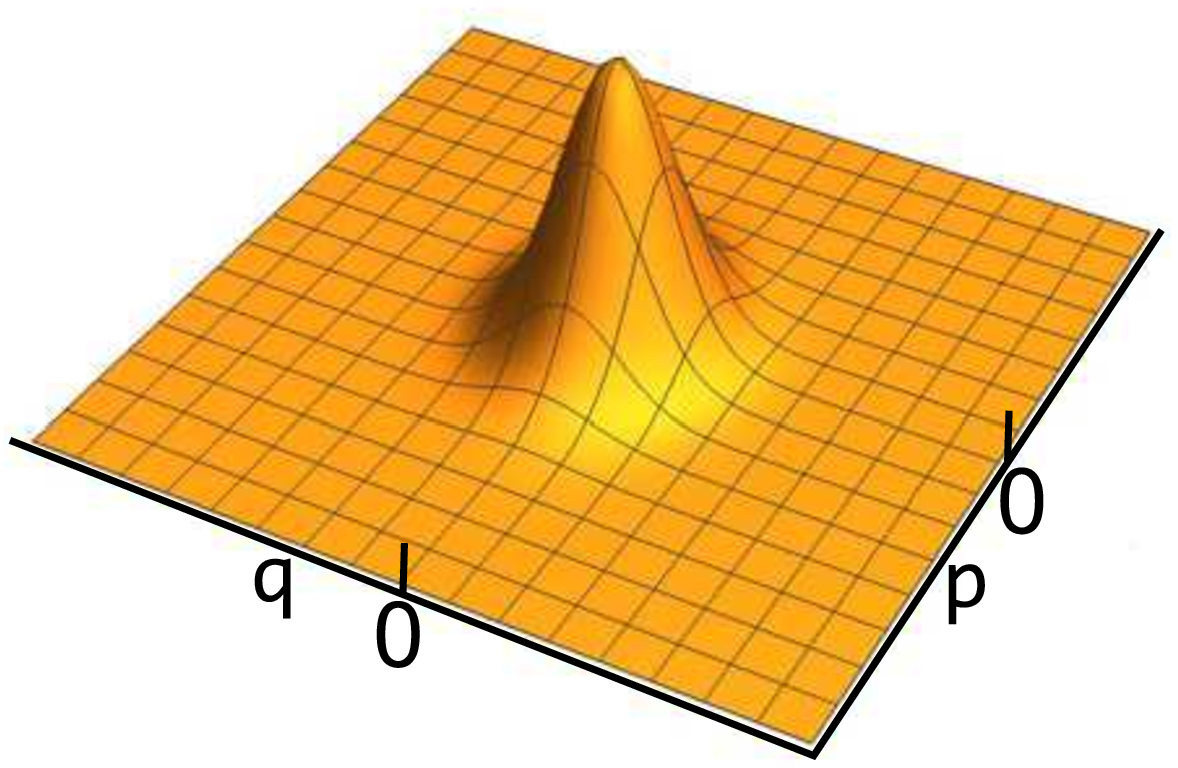}}
\subfloat[e]{\includegraphics[scale=0.2]{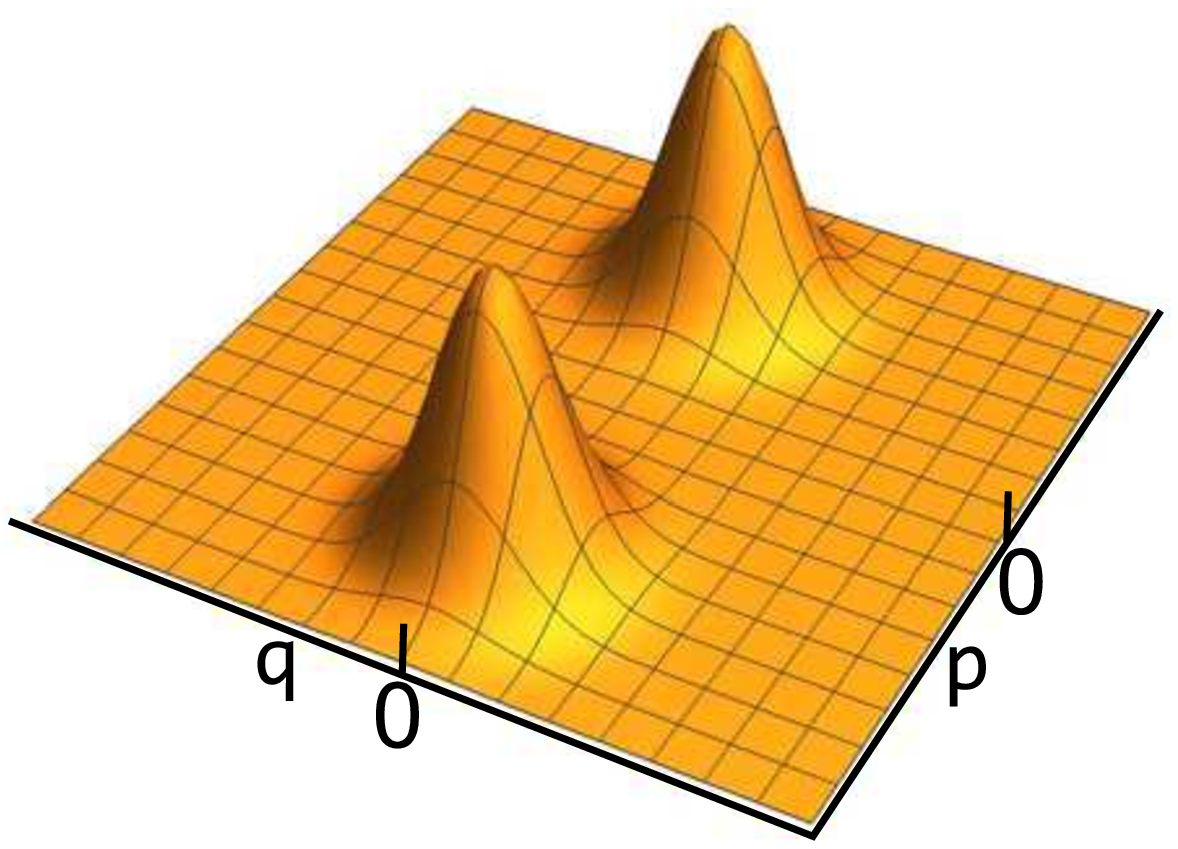}}
\caption{Evolution of Husimi Q--functions, with $q\sqrt{\omega/2},p/\sqrt{2\omega}$ normalized coordinates of photons, for the routes in Fig.\ref{fig:PDG} at $\omega = E_J = 1$, $S = 2$: a) Crossover from normal to subradiant phase; trajectory 1 at $\epsilon = 0.5$; b) Crossover from normal to superradiant broken parity phase; trajectory 2 at $\epsilon = 0.04$, $\alpha = 0.41$; c) Crossover from normal to superradiant precursor, parity symmetric phase, trajectory 3 at $\epsilon = -0.3$. See text for details. Husimi Q--function for photon subsystem at $S=3/2$, $\epsilon = 0.2$: d) $g=0$; e) $g=10$.}
\label{fig:HQ}
\end{figure}

Both analytical and numerical approaches were used to obtain the phase diagram. Numerically, the different phases can be distinguished using the dependence of the spin subsystem entropy on the coupling constant \cite{Rabl}. This entropy is defined as:
\begin{equation}
\mathcal{S} = -\Tr \rho_S \log \rho_S,
\end{equation} 
where $\rho_S$ is the reduced spin density matrix defined in (\ref{eq:rhoS}). For fixed $\varepsilon$, we consider coupling constant $g_{cr}$, at which the entropy achieves its maximal value, as belonging to a crossover 'line' $g_{cr}(\varepsilon)$ for $\varepsilon>0$, or to a transition line $g(\varepsilon)$ into superradiant precursor phase for  $\varepsilon<0$, see corresponding solid lines in Fig. \ref{fig:PDG}. An example of the calculated $S(g)$ dependencies is shown in Fig. \ref{fig:Entropy}. 
While the maximum is rather wide for $\varepsilon>0$, it is rather narrow in the $\varepsilon<0$ case. Corresponding evolution histories of the Husimi Q--functions in these two cases are calculated and presented in Fig. \ref{fig:HQ} a,c.

For analytical calculations at small finite $\varepsilon$, we apply the same method of not commuting limits, $g \rightarrow \infty, \alpha\rightarrow 0$  \cite{Wezel}, used above for $\varepsilon=0$ case. Considering now finite inter-dipole interaction term, $\propto \varepsilon g^2\hat{S}_y^2$ in the Hamiltonian (\ref{eq:H2}), it is straightforward to modify expression in (\ref{E2d}) by adding $\varepsilon$-dependent interaction term, leading to the following ground state energy up to the second order perturbation in $E_J$:

\begin{equation}\label{E2de}
E_0=E^{(0)}_{\alpha,\varepsilon} +E^{(2)}_{\alpha,\varepsilon}= - \dfrac{\alpha^2}{2}-g \alpha \sigma_y +\frac{E_J^2}{2g^2}\left(\sigma_y^2-S(S+1)\right)+\dfrac{\varepsilon g^2}{2}\sigma_y^2.
\end{equation}
\noindent First, we mention, that without the symmetry-breaking terms, i.e. at $\alpha=0$, corresponding to l.h.s. order of taking limits in (\ref{eq:lims}), this expression was considered previously \cite{Rabl_2016}. The authors found proliferation of the subradiant domain below the $\epsilon=0$ axis into the region $-E_J^2/g^4<\varepsilon<0$, where (\ref{E2de}) would be concave with respect to variable $\sigma_y $, thus indicating subradiant behaviour. Our analysis shows that a finite symmetry breaking perturbation $\alpha$ alters the latter conclusion. Namely, minimize $E_0$ in (\ref{E2de}) with respect to $\sigma_y$  at finite $\alpha$ :
\begin{equation}\label{varalpm}
\frac{\partial E_0 }{\partial\sigma }=0;\; \Rightarrow \alpha_m=(E_J^2/g^3+\varepsilon g)\sigma_y.
\end{equation}
\noindent
Now, substitute the resulting expression for $\alpha_m$ back into (\ref{E2de}) and find the minimal energy as function of $\sigma_y(\alpha)$:
\begin{equation}\label{Econvexe}
E_{\varepsilon,rhs}=E^{(0)}_{\alpha_m,\varepsilon} +E^{(2)}_{\alpha_m,\varepsilon}= - \frac{E_J^2}{2g^2} \left(\sigma_y^2+S(S+1)\right)-\dfrac{\varepsilon g^2}{2}\sigma_y^2 +O(E_J^4/g^{6}).
\end{equation}
\noindent Thus, the curvature of the free energy with respect to projection $\sigma_y$ has changed its sign and became convex for $\varepsilon>0$, indicating superradiant instability in the system. In order to follow  the r.h.s order of limits indicated in (\ref{eq:lims}), i.e. the limit $\alpha\rightarrow 0$ follows the limit $g \rightarrow \infty$, the following inequality has to be fulfilled:

\begin{equation}\label{alpde}
\frac{\partial |\alpha_m| }{\partial g }<0; \; \Rightarrow 0<\varepsilon<3E_J^2/g^4.
\end{equation}

\noindent The upper boundary of the domain of the phase diagram, where (\ref{alpde}) is fulfilled, is indicated in Fig. \ref{fig:PDG} (see dotted line). 

For $\varepsilon\equiv -|\varepsilon| <0$ the ground state energy in (\ref{Econvexe}) turns into: 
\begin{equation}\label{Econvexe1}
E_{\varepsilon,rhs}=E^{(0)}_{\alpha_m,\varepsilon} +E^{(2)}_{\alpha_m,\varepsilon}= - \frac{E_J^2}{2g^2} \left(\sigma_y^2+S(S+1)\right)+\dfrac{|\varepsilon| g^2}{2}\sigma_y^2 +O(E_J^4/g^{6});\;\varepsilon\equiv -|\varepsilon|.
\end{equation}
\noindent 
Thus, the ground state energy preserves convex curvature with respect to projection $\sigma_y$, provided:
\begin{equation}\label{convexee}
 -E_J^2/g^4<\varepsilon<0.
\end{equation}
\noindent The above inequality also guarantees compliance with the condition of the r.h.s order of limits in (\ref{eq:lims}), i.e.:
\begin{equation}\label{convexe2}
\alpha_m=(E_J^2/g^4+\varepsilon )g\sigma_y =\mbox{O}(E_J^2/g^3)\rightarrow 0|_{g\rightarrow \infty}.
\end{equation}  
Hence, the part of the phase diagram under the $g^2$-axis, where $\varepsilon<0$, delimited from the bottom by inequality (\ref{convexee}), is also unstable with respect to superradiant transition under an external perturbation $\sim \alpha$, that vanishes with the growth of coupling constant $g$ at fixed size $N$ of the TLS system, see dashed line in Fig.\ref{fig:PDG}. Below we present our results of the exploration of the described above phase diagram using the method of Husimi Q-functions \cite{Husimi}, see Appendix for definitions. Obtained images of the Q-functions, Fig.\ref{fig:HQ}, permit one to follow directly the evolution of the state of the finite $N$ pseudospin-photon system along the three different routes with increasing coupling strength $g^2$ within the three domains of the phase diagram in Fig. \ref{fig:PDG}. The arrows along the trajectories connect the points (filled circles),  for which numerical calculations were performed. We will analyse spin and photon Husimi Q--functions, that are calculated using density matrix of the ground state. The latter state is distinguished by the minimal eigenvalue of the Hamiltonian matrix.  

\subsection{Husimi Q-functions in subradiant phase $\varepsilon >0$}
Travel along the trajectory $1$ in Fig.\ref{fig:PDG} leads to the set of Husimi Q--functions presented in Fig.\ref{fig:HQ}, a.  At zero coupling $g$ the energy is minimal, when pseudospin-projection on $z$--axis has maximal  value. Simultaneously, the probability weight of the  photon subsystem is peaked around $\{0,0\}$ in the $\{q,p\}$ plane. In the opposite case of strong coupling $g$, since $\varepsilon >0$, the spin projection on $y$--axis tends to be minimal and pseudospin Q-function acquires "bagel"-shape. Simultaneously, since pseudospin is integer, $S=2$, the  photon subsystem remains peaked around $\{0,0\}$ in the $\{q,p\}$ plane, as is obvious e.g. from Fig. \ref{fig:BO_pot_up} for integer spin $S$. The crossover from "normal" to "subradiant" phase of the spin subsystem is clear from the middle plot in Fig.\ref{fig:HQ},a of Husimi pseudospin Q--function. For the case of odd $N$ (half-integer pseudospin $S$), the evolution for the photon subsystem Q--function from one peak around $\{0,0\}$ in the $\{q,p\}$ plane at $g=0$, to the twin peaks $p=\pm g/2$, symmetric around $p=0$, in the limit of strong coupling $g$, is presented in the separate sets in Fig.\ref{fig:HQ}d, e. This situation is explained in the Fig. \ref{fig:BO_pot_up} for half-integer spin $S$. The distance between the peaks, though increases with coupling strength $g$, is independent of $S=N/2$, i.e. independent of the 'size' $N$ of the TLS array. 

\subsection{Husimi Q-functions in symmetry broken phase for small $|\varepsilon|$}
Evolution along the trajectory $2$ in Fig.\ref{fig:PDG} leads to the set of Husimi Q--functions presented in Fig.\ref{fig:HQ}, b. The qualitative difference
of the  trajectory $2$ from the trajectories 1 and 3, is a broken parity phase, as indicated in Fig. \ref{fig:PDG} and discussed in subsection "Superradiant instability at small $|\varepsilon|$" above. The remarkable feature seen in pseudospin Husimi Q--function evolution in Fig.\ref{fig:HQ}, b, is rotation by $90^0$ degrees angle in the $\{S_y,S_z\}$ plane of the total pseudospin, starting from $g=0$ and saturating at $g\rightarrow\infty $, as was predicted earlier in \cite{Mukhin} in $N>>1$ case. Simultaneously, the photon subsystem Q--function migrates from $\{0,0\}$ to e.g. $\{0, -gS\}$ in the $\{q,p\}$ plane (an option $\{0, gS\}$ is also possible as an alternative choice of broken symmetry state), as is explained in subsection "Superradiant instability at $\varepsilon = 0$", Eq. (\ref{avp}).
\subsection{Husimi Q-functions in 'superradiant' phase $\varepsilon <0$}
Finally, evolution along the trajectory $3$ in Fig.\ref{fig:PDG} leads to the set of Husimi Q--functions presented in Fig.\ref{fig:HQ}, c. Unlike in subradiant phase, the pseudospin Q--function becomes symmetrically dumbbell-shaped, when the coupling strength $g^2$ increases. This is explained in 
Fig.\ref{fig:BO_pot_down}. The latter demonstrates, that $S_y\rightarrow \pm S$ in the ground state of a finite $S=N/2$ system, that nevertheless, remains parity symmetric, until the symmetry is broken at the phase transition in $g^2\rightarrow \infty$ limit, and only half of the dumbbell would survive, as explained in subsection "Superradiant instability at small $|\varepsilon|$" above, and also shown in Fig. \ref{fig:HQ}, b. Simultaneously, photon subsystem Q--function evolves from the one  peak around $\{0,0\}$ in the $\{q,p\}$ plane at $g=0$, to the twin peaks separated proportionally to the TLS 'size':  $p=\pm gS$, in the limit of strong coupling $g$, in contradistinction with the $subradiant$ case of the half-integer pseudospin system $S=N/2$, where the twin peaks  move away from one another according to relation $p=\pm g/2$, i.e. independent of the 'size' of the TLS. Then, at the superradiant phase transition, the system 'jumps' from Q-shapes in Fig.\ref{fig:HQ}, c to the ones in Fig.\ref{fig:HQ}, b.

\section{Variational approach at $\varepsilon = 0$}\label{sec:Variation}
In the case of $\varepsilon = 0$  we have also applied a variational function method, to study spin configuration in the ground state of the Hamiltonian in Eq. (\ref{Dickeini}), with $\varepsilon = 0$. The test function is:
\begin{equation}\label{eq:test_func}
\psi(a, b) = \sum_{\sigma_y = -S}^S \sum_{n = 0}^\infty \underbrace{\qty(\frac{a}{(n^2 + \sigma_y^2 + 1)^2}+ \frac{b}{(n^2 + (\sigma_y-S)^2 + 1)^2})}_{c_{n,\sigma_y}}\ket{n_{g\sigma_y}, \sigma_y}.
\end{equation}
It is asymmetric with respect to transformation $\sigma_y \rightarrow -\sigma_y$, so one can study the broken symmetry phase. The latter is reached in the thermodynamic limit, $N \gg 1$,  via a superradiant first order phase transition, found previously in the $\varepsilon = 0$ case \cite{Mukhin}, or alternatively, is attained in the strong coupling  limit $g \gg 1$ for a finite system with $N\simeq 1$. For numerical calculation the infinite sum over $n$ was truncated and a sufficient accuracy was achieved at $n \sim 100$. Coefficient $b$ can be expressed as function of $a$, using normalisation condition:
\begin{equation}\label{eq:norm}
\sum_{\sigma_y = -S}^S \sum_{n = 0}^\infty \qty(\frac{a}{(n^2 + \sigma_y^2 + 1)^2}+ \frac{b}{(n^2 + (\sigma_y-S)^2 + 1)^2})^2 = 1.
\end{equation}  
Hence, we obtain test function $\psi(a,b(a)) = \psi(a)$, that depends on a single parameter $a$. The equation (\ref{eq:norm}) is solved numerically and real solutions exist in the range $a \in [0, \simeq 0.9]$. Ground state energy and wave function are found by minimizing $E(a) = \ev{\hat H}{\psi(a)}$ as function of $a$, where $\hat H$ is Hamiltonian (\ref{Dickeini}) with $\varepsilon=0$, and coefficients $c_{n,\sigma_y}$ are defined in Eq. (\ref{eq:test_func}):
\begin{equation}
\begin{aligned}
E(a) &= \sum_{n = 0}^\infty \sum_{\sigma_y = -S}^S c_{n,\sigma_y}^2\omega n -\\
& - \frac{E_J}{2}\sum_{n,m = 0}^\infty \sum_{\sigma_y = -S}^S c_{n,\sigma_y} \left(c_{m,\sigma_y+1} \Omega_{n,m}^{\sigma_y, \sigma_y + 1}\sqrt{S(S+1) - \sigma_y(\sigma_y + 1)} + \right.\\
& \left. + c_{m,\sigma_y - 1} \Omega_{n,m}^{\sigma_y, \sigma_y - 1}\sqrt{S(S+1) - \sigma_y(\sigma_y - 1)}\right).
\end{aligned}
\end{equation}
Here $\Omega_{n,m}^{\sigma_y, \sigma'_y} = \braket{n_{g\sigma_y}}{m_{g\sigma_y'}}$ is the overlap of photon harmonic oscillator shifted equilibrium states \cite{Waldenstrom}. Given $E(a_0)$ is minimal at $a_0$, we can decide what is the projection of the spin on y--axis in the ground state with the wave function $\psi(a_0)$: if $a_0 = 0$ the second term in (\ref{eq:test_func}) is dominating and $\ev{S_y} = S$, while $\ev{S_y} \rightarrow 0$ when $a_0 \rightarrow 1$.

\begin{figure}[h]
\subfloat[$S=\frac{3}{2}$]{\includegraphics[scale=0.6]{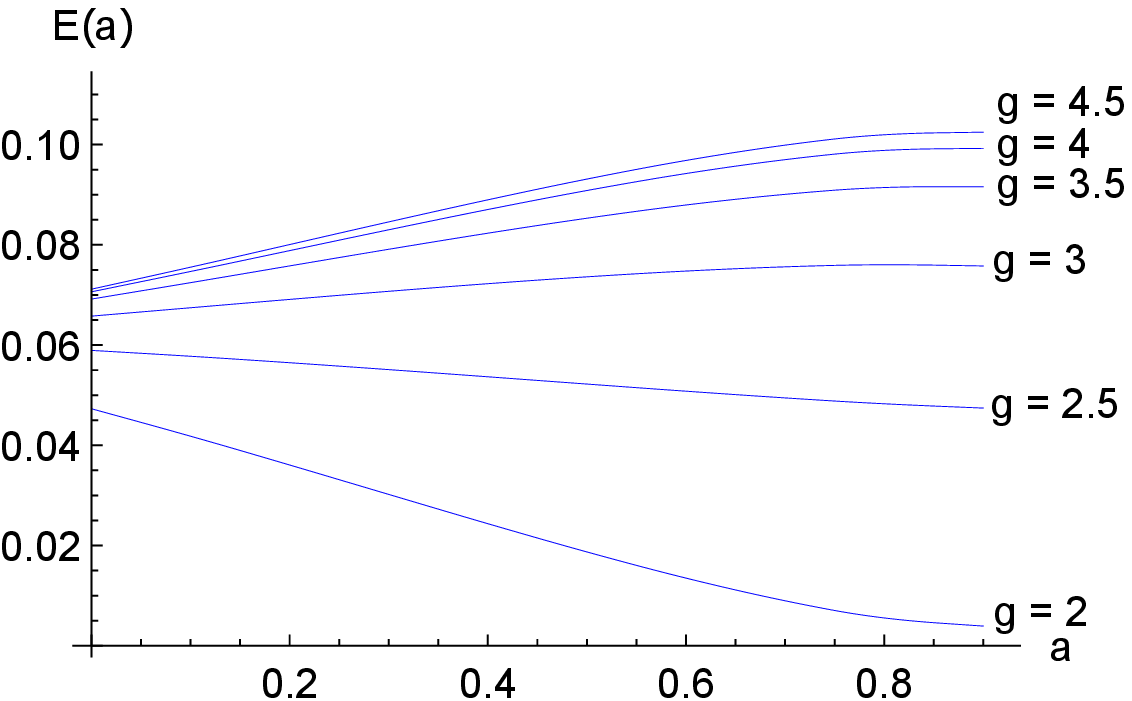}}\hspace{1cm}
\subfloat[$S=2$]{\includegraphics[scale=0.6]{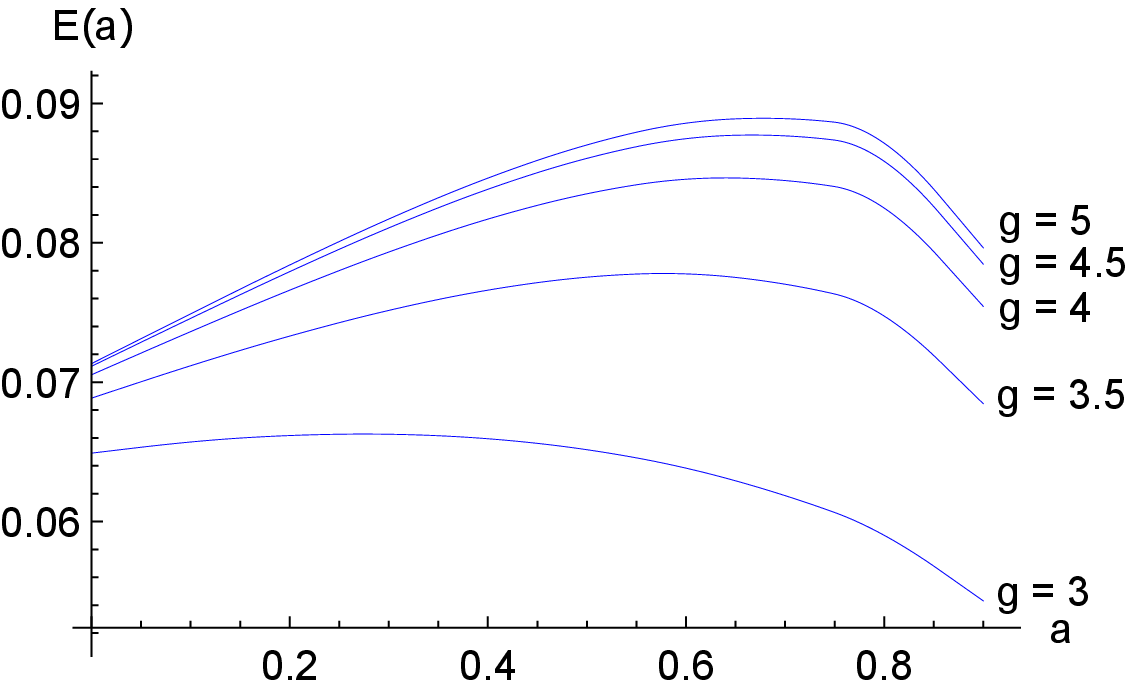}}
\caption{Dependence of the ground state energy on parameter $a$ for $g$ in the range from: $2$ to $4.5$ (left panel), and from $3$ to $5$ (right panel). The coupling constant grows from lowest curve to the highest one. The summation over number of photons was truncated at n $\simeq100$. }
\label{fig:Variation}
\end{figure}

\begin{figure}[h]
\center\includegraphics[scale=0.6]{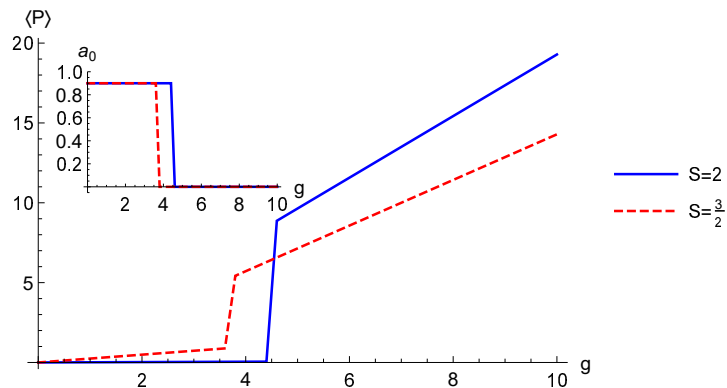}
\caption{Dependence of photon oscillator average  momentum absolute value on coupling constant $g$. The inset demonstrates dependence of $a_0$, which minimizes the energy, on coupling constant $g$.}
\label{fig:P(g)}\end{figure}

The result now does not depend on spin being integer or half--integer. In Fig.\ref{fig:Variation} and inset of Fig. \ref{fig:P(g)} one can see that the minimum of $E(a)$ changes its location from $a_0 > 0$ to $a_0 = 0$ when coupling constant $g$ increases. It means that at $g>g_c$, where $a_0=0$, the system behaves as it does at $\varepsilon < 0$ case, possessing maximal absolute value of the pseudospin  projection along the $y$ -- axis, and manifestly broken parity symmetry of the wave-function (\ref{eq:test_func}) .
\noindent 
\begin{figure}[h]
\center\includegraphics[scale=0.7]{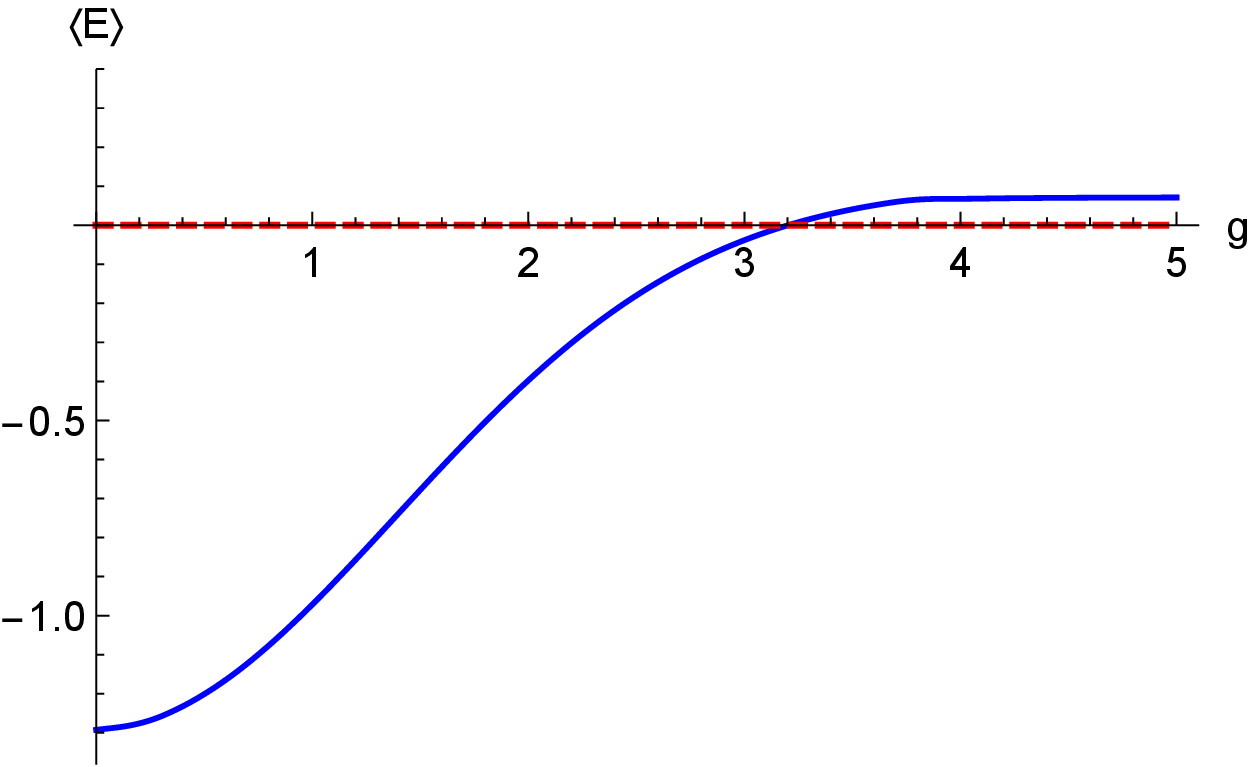}
\caption{Dependence of ground state energy on coupling constant, calculated with the use of the test function Eq.(\ref{eq:test_func}). The dashed line denotes asymptotic value at strong coupling.}
\label{fig:E0(g)}
\end{figure}

We calculate average of momentum operator $\ev{\hat{p}}{\psi(a_0)}=-g\ev{\hat{S}_y}{\psi(a_0)}$ in the ground state as function of coupling constant $g$, using the variational wave function $\psi(a_0)$  found above. As a result, we found, that there is a jump from $\ev{p}=0$ to $\ev{p} \neq 0$ at the transition point when $\ev{S_y}$ becomes non-zero. This jump signifies emergence of coherent photonic state in the system (compare \cite{Mukhin}).

We conclude that at $\varepsilon = 0$ and $g\rightarrow \infty$ the system behaves like in $\varepsilon < 0$ case, i.e. the ground state wave-function breaks the parity symmetry and concentrates at the maximal absolute value of projection of the pseudospin on $y$--axis. This is in agreement with analysis presented in \cite{Mukhin} and with the results presented above and summarised in the Figs. \ref{fig:PDG}, \ref{fig:HQ}. To check an accuracy of the chosen test function we calculate dependence of the ground state energy on coupling constant and compare it with known asymptotics at ultrastrong coupling --- $E_0(g \rightarrow \infty) = 0$ (counted from $\omega/2$). The result is presented in Fig. \ref{fig:E0(g)}. As expected, the test-function energy is slightly larger, than the exact one.

\section{Conclusion}
The phase diagram of finite array of $N< \infty$ two-level systems (TLS), coupled to a photon (harmonic oscillator) mode and described by extended Dicke model Hamiltonian is investigated. An infinitesimal parity symmetry breaking field, proportional to a small parameter $\alpha\rightarrow 0$, is added to the Hamiltonian. A check of commutation of the impacts on the ground state wave function of the $g\rightarrow \infty $ and $\alpha\rightarrow 0$ limits is made. Obtained analytical results, as well as numerical calculations of the Husimi Q-functions, indicate that the system is unstable with respect to superradiant transition inside an area of the phase diagram, where impacts of the vanishing external perturbation $\sim \alpha$, and of increasing photon-TLS coupling $\propto g$, do not commute.  In a fixed 'size' $N$ TLS, $g \rightarrow \infty$ plays the role of thermodynamic limit. The calculated evolution of the Husimi Q-functions in the subradiant area of the phase diagram provides manifestation of the nature of the crossover between normal and subradiant phases. The crossover line $g(\varepsilon)$ is found by locating maxima of the pseudo-spin subsystem entropy, that appear with increasing coupling $g$. Behaviour of photonic  Husimi Q-functions differs for odd and even numbers $N$ of TLS, while pseudospin Q-functions evolve similarly in both cases. Evolution  in subradiant phase of the distance between twin peaks of the photonic Husimi Q-functions in harmonic oscillator phase plane $\{q,p\}$ depends on the parity of the TLS number $N$, but not on its value. This is in contradistinction with the superradiant precursor phase, with finite $N$ and $g$, where the distance is proportional to $gN$.  It is remarkable, that evolution of the pseudospin Husimi Q--function of finite $N$ TLS, with added infinitesimal parity symmetry breaking field, exhibits rotation of the total pseudospin by $90^0$ angle in the $\{S_y,S_z\}$ plane, starting from $g=0$ and saturating at $g\rightarrow\infty $, as was predicted earlier in \cite{Mukhin} in $N>>1$ case.


\section*{Acknowledgement}
S.I. acknowledges warm hospitality and useful discussions with prof. C.W.J. Beenakker and members of his group during his stay in Lorentz Institute. The authors acknowledge the financial support of the Ministry of Science and Higher Education of the Russian Federation in the framework of Increase Competitiveness Program of NUST MISiS, Grant No. K2-2020-001. The work of S.S.S. was also supported in part by Russian Science Foundation grant 18-12-00438. 

\appendix
\section{Husimi Q--functions}
We briefly describe below an essence of the Husimi Q-functions method \cite{Husimi}. 
Given an eigenstate $\ket{\Psi}$ of the whole system one can obtain reduced density matrix of a constituting subsystem by a partial trace operation:
\begin{align}
\rho_\text{Ph} &= \operatorname{Tr}_\text{S}\ketbra{\Psi}\\
\rho_\text{S} &= \operatorname{Tr}_\text{Ph}\ketbra{\Psi}\label{eq:rhoS}
\end{align}
where $\operatorname{Tr}_{\text{S},\text{Ph}}$ denotes partial trace over pseudospin or photon subsystem respectively. Husimi Q--function is then the reduced density matrix averaged over a coherent state, i.e. if the coherent state is parametrised by a set of variables $\alpha_1, \ldots, \alpha_n$
\begin{equation}
Q(\alpha_1, \ldots, \alpha_n) = \bra{\alpha_1, \ldots, \alpha_n}\rho\ket{\alpha_1, \ldots, \alpha_n}.
\end{equation} 
Its value gives the probability of the subsystem being in the state $\ket{\alpha_1, \ldots, \alpha_n}$. 

\subsection{Calculation of reduced density matrix}
The numerical procedure for calculation of reduced density matrix is described in this section. First, the Hamiltonian matrix is obtained by expanding the Hamiltonian in basis of states $\ket{n, \sigma_y}$. Sufficiently large number of photon states $n \sim 100$ should be used in order to achieve desired accuracy. The matrix is arranged in such a way, that it consists of $(2S + 1) \times (2S + 1)$ blocks. Each block corresponds to some fixed photon state and the spin states changing from $-S$ to $S$, i.e. the order of the rows and columns is $\ket{0,-S}, \ldots, \ket{0, S}, \ket{1, -S},\ldots \ket{1, -S}, \ldots$. The ground state wave function is an eigenvector of the Hamiltonian matrix and ground state density matrix is calculated by multiplication of the eigenvector on its conjugate. Given mentioned above arrangement, $\Tr_\text{S}$ means taking trace in each spin-block of density matrix, and $\Tr_\text{Ph}$ means adding up diagonal blocks.

\subsection{Spin subsystem}
Spin coherent state is parametrised by spherical angles $\phi$ and $\theta$ and can be obtained by an action of rotation operator \cite{Spin_squeezing_Ma}
\begin{equation}
\hat R(\theta, \phi) = e^{i \theta (\hat S_z \sin \phi - \hat S_y \cos\phi)}
\end{equation}
on the eigenstate of $\hat S_z$, i.e. $\ket{\theta,\phi} = \hat R(\theta,\phi)\ket{\sigma_z = S}$. Husimi Q--function is given then by 
\begin{equation}
Q_S(\theta, \phi) = \bra{\theta,\phi}\rho_S\ket{\theta,\phi}.
\end{equation}

\subsection{Photon subsystem}
One can write Q--function for photons as well:
\begin{equation}
Q_{Ph}(p, q) = \bra{p,q}\rho_{Ph}\ket{p,q},
\end{equation} 
where $\rho_{Ph}$ is the photon density matrix, $\ket{p,q}$ is a coherent photon state with $\ev{\hat p} = p$ and $\ev{\hat q} = q$. This function demonstrates probability density of photon having coordinates $p$ and $q$ in the phase space. In the normal phase in Fig. \ref{fig:PDG} the distribution is peaked at $p=0,q=0$, meaning there is no photon condensate. 

\bibliographystyle{ieeetr}

\bibliography{biblio} 

\begin{thebibliography}{10}

\bibitem{Dicke}
R.~H. Dicke, ``Coherence in spontaneous radiation processes,'' {\em Phys.
  Rev.}, vol.~93, pp.~99--110, Jan 1954.

\bibitem{Garraway}
B.~M. Garraway, ``The {D}icke model in quantum optics: {D}icke model
  revisited,'' {\em Philosophical Transactions of the Royal Society A:
  Mathematical, Physical and Engineering Sciences}, vol.~369, no.~1939,
  pp.~1137--1155, 2011.

\bibitem{Brandes}
C.~Emary and T.~Brandes, ``Chaos and the quantum phase transition in the
  {D}icke model,'' {\em Physical review. E, Statistical, nonlinear, and soft
  matter physics}, vol.~67, p.~066203, 06 2003.

\bibitem{Hepp}
K.~Hepp and E.~H. Lieb, ``On the superradiant phase transition for molecules in
  a quantized radiation field: the {D}icke maser model,'' {\em Annals of
  Physics}, vol.~76, no.~2, pp.~360 -- 404, 1973.

\bibitem{Saidi}
W.~Saidi and D.~Stroud, ``Eigenstates of a small {J}osephson junction coupled
  to a resonant cavity,'' {\em Physical Review B}, vol.~65, p.~014512, 2001.

\bibitem{Rabl}
D.~De~Bernardis, T.~Jaako, and P.~Rabl, ``Cavity quantum electrodynamics in the
  nonperturbative regime,'' {\em Phys. Rev. A}, vol.~97, p.~043820, Apr 2018.

\bibitem{Rabl_2016}
T.~Jaako, Z.-L. Xiang, J.~J. Garcia-Ripoll, and P.~Rabl, ``Ultrastrong-coupling
  phenomena beyond the {D}icke model,'' {\em Phys. Rev. A}, vol.~94, p.~033850,
  Sep 2016.

\bibitem{Mukhin}
S.~I. Mukhin and N.~V. Gnezdilov, ``First-order dipolar phase transition in the
  {D}icke model with infinitely coordinated frustrating interaction,'' {\em
  Phys. Rev. A}, vol.~97, p.~053809, May 2018.

\bibitem{Rzazewski}
K.~Rzazewski, K.~Wodkiewicz, and W.~Zakowicz, ``Phase transitions, two-level
  atoms, and the ${A}^{2}$ term,'' {\em Phys. Rev. Lett.}, vol.~35,
  pp.~432--434, Aug 1975.

\bibitem{Keeling}
J.~Keeling, ``{C}oulomb interactions, gauge invariance, and phase transitions
  of the {D}icke model,'' {\em Journal of Physics: Condensed Matter}, vol.~19,
  p.~295213, jun 2007.

\bibitem{Stokes}
A.~Stokes and A.~Nazir, ``Uniqueness of the phase transition in many-dipole
  systems,'' {\em arXiv:1905.10697}, 05 2019.

\bibitem{Bialynicki-Birula}
I.~Bialynicki-Birula and K.~Rza\ifmmode \mbox{\c{}}\else
  \c{}\fi{}\ifmmode~\dot{z}\else \.{z}\fi{}ewski, ``No-go theorem concerning
  the superradiant phase transition in atomic systems,'' {\em Phys. Rev. A},
  vol.~19, pp.~301--303, Jan 1979.

\bibitem{Ashhab}
S.~Ashhab, ``Superradiance transition in a system with a single qubit and a
  single oscillator,'' {\em Phys. Rev. A}, vol.~87, p.~013826, Jan 2013.

\bibitem{IMAI20183333}
R.~Imai and Y.~Yamanaka, ``Stability of symmetry breaking states in finite-size
  {D}icke model with photon leakage,'' {\em Physics Letters A}, vol.~382,
  no.~46, pp.~3333 -- 3338, 2018.

\bibitem{Shen_2017}
L.-T. Shen, Z.-B. Yang, H.-Z. Wu, and S.-B. Zheng, ``Quantum phase transition
  and quench dynamics in the anisotropic {R}abi model,'' {\em Phys. Rev. A},
  vol.~95, p.~013819, Jan 2017.

\bibitem{Shen_2020}
L.~Shen, Z.~Shi, Z.~Yang, H.~Wu, Z.~Zhong, and S.~Zheng, ``A similarity of
  quantum phase transition and quench dynamics in the {D}icke model beyond the
  thermodynamic limit,'' {\em EPJ Quantum Technology}, vol.~7, 12 2020.

\bibitem{Hwang}
M.-J. Hwang, R.~Puebla, and M.~B. Plenio, ``Quantum phase transition and
  universal dynamics in the {R}abi model,'' {\em Phys. Rev. Lett.}, vol.~115,
  p.~180404, Oct 2015.

\bibitem{Wezel}
J.~{van Wezel} and J.~{van den Brink}, ``Spontaneous symmetry breaking in
  quantum mechanics,'' {\em American Journal of Physics}, vol.~75, p.~635, 3
  2007.

\bibitem{Aron}
A.~J. Beekman, L.~Rademaker, and J.~van Wezel, ``{An Introduction to
  Spontaneous Symmetry Breaking},'' {\em SciPost Phys. Lect. Notes}, p.~11,
  2019.

\bibitem{Cohn}
J.~Cohn, A.~Safavi-Naini, R.~J. Lewis-Swan, J.~G. Bohnet, M.~Gärttner, K.~A.
  Gilmore, J.~E. Jordan, A.~M. Rey, J.~J. Bollinger, and J.~K. Freericks,
  ``Bang-bang shortcut to adiabaticity in the {D}icke model as realized in a
  penning trap experiment,'' {\em New Journal of Physics}, vol.~20, p.~055013,
  may 2018.

\bibitem{Bastarrachea_lattice}
M.~A. Bastarrachea-Magnani and J.~G. Hirsch, ``Peres lattices and chaos in the
  {D}icke model,'' {\em Journal of Physics: Conference Series}, vol.~512,
  p.~012004, may 2014.

\bibitem{Husimi}
K.~HUSIMI, ``Some formal properties of the density matrix,'' {\em Proceedings
  of the Physico-Mathematical Society of Japan. 3rd Series}, vol.~22, no.~4,
  pp.~264--314, 1940.

\bibitem{Waldenstrom}
S.~Waldenstrøm and K.~Naqvi, ``The overlap integrals of two
  harmonic-oscillator wavefunctions: some remarks on originals and
  reproductions,'' {\em Chemical Physics Letters}, vol.~85, no.~5, pp.~581 --
  584, 1982.

\bibitem{Spin_squeezing_Ma}
J.~Ma, X.~Wang, C.~Sun, and F.~Nori, ``Quantum spin squeezing,'' {\em Physics
  Reports}, vol.~509, no.~2, pp.~89 -- 165, 2011.

\end{thebibliography}
\end{document}